\documentclass[aps,prl,twocolumn,a4paper,10pt]{revtex4-1}

\usepackage[T1]{fontenc}    
\usepackage[english]{babel}   
\usepackage[utf8]{inputenc} 
\usepackage{times}

\usepackage{amsmath}      
\usepackage{amssymb}      
\usepackage{bbm}        
\usepackage{mathtools}
\usepackage{simpler-wick}
\usetikzlibrary{tikzmark}
\DeclarePairedDelimiterX\Dirbraket[3]{\langle}{\rangle}%
{#1\,\delimsize\vert\,\mathopen{}#2\,\delimsize\vert\,\mathopen{}#3}

\usepackage[colorlinks=true, a4paper=true, pdfstartview=FitV,linkcolor=blue, citecolor=blue, urlcolor=blue]{hyperref}

\usepackage{graphicx}     
\usepackage{graphics}     
\DeclareGraphicsExtensions{.pdf}
\usepackage{color}

\newcommand{\bea}{\begin{eqnarray}}
\newcommand{\eea}{\end{eqnarray}}

\newcommand{\mbf}{\mathbf}

\newcommand{\br}{\mathbf{r}}

\definecolor{darkGreen}{rgb}{0.0, 0.5, 0.0}

\begin{document}

\title{Ab initio description of magnetic polarons in a Mott insulator}

\author{Emil Blomquist$^{1}$ and Johan~Carlstr\"om$^{2}$ }
\affiliation{$^{1}$Department of Physics, Royal Institute of Technology, Stockholm, SE-106 91, Sweden}
\affiliation{$^{2}$Department of Physics, Stockholm University, 106 91 Stockholm, Sweden}
\date{\today}

\begin{abstract}
Polarons are among the most elementary quasiparticles of interacting quantum matter, consisting of a charge carrier dressed by an excited background \cite{Polaron}. In Mott insulators, they take the form of a dopant surrounded by a distorted spin-background and are expected to dictate effective mass, transport properties and interactions between carriers  \cite{PhysRevB.39.6880,PhysRevLett.116.247202,PhysRevB.37.1597}. 
Despite the fundamental importance of polarons for the electronic structure of strongly correlated systems, access to their internal structure was only recently realized in experiments \cite{Koepsell2019}, while theoretical results are still lacking due to the sign problem. Here we report unbiased high-precision data obtained from worm-algorithm Monte Carlo that reveal the real-space structure of a polaron in the $t$-$J$ model deep inside the region where the sign problem becomes significant. These results are directly comparable to recent quantum gas microscopy experiments, but give access to significantly lower temperatures.
\end{abstract}
\maketitle

The doped Mott insulator with charge carriers that propagate on an anti-ferromagnetic background has been identified as the paradigmatic model of the high-temperature superconductors \cite{RevModPhys.78.17}. In this scenario, the dopants and the environment form composite polarons as a result of competition between the kinetic energy and super-exchange processes in the background \cite{PhysRev.118.141}.
The carriers can lower their energy by delocalizing, but this is incompatible with anti-ferromagnetism as their motion disrupts the magnetic order.
Retaining anti-ferromagnetic correlations minimizes the exchange energy, but this effectively leads to narrowing of the band due to a finite mean-free path \cite{brink}--delocalization of the carrier within a cloud of altered spin-correlations results as a compromise.

Since the polaron is the elementary quasiparticle of the doped Mott insulator, it is also the starting point for understanding its electronic structure, including the enigmatic mechanism of high-temperature superconductivity. While still a matter of active research, the premise that the perturbed spin-background mediates attraction between carriers is considered to be central to the explanation of pairing \cite{ANDERSON1196,PhysRevB.39.11663}. Attraction between carriers in a Mott insulator is also indicated by density matrix renormalization group (DMRG) calculations  \cite{PhysRevB.55.6504}. 

Currently, to the best of our knowledge, no approximation-free theoretical results without systematic bias exist for the internal structure of the polaron. Exact diagonalization (ED) can resolve spin-correlations in the vicinity of the carrier, but the restriction to small system sizes means that the polaron is distorted by boundary pressure (see also supplementary information). DMRG allows addressing larger systems than ED but is strictly speaking an approximative technique, albeit potentially very accurate \cite{PhysRevB.48.10345}. In macroscopic systems, the fermionic sign problem prevents the effective use of conventional quantum Monte Carlo techniques \cite{PhysRevB.41.9301}.

Until recently, detailed information about polarons was equally unavailable in experiments, as electronic systems are generally not accessible to the required precision.
With the advent of quantum gas microscopy, this has changed completely,  and imaging of entangled quantum many-body states is now possible at the level of single-site resolution \cite{Gross995}. The experimental realization of strongly correlated systems in ultra-cold atomic gases has at this point reached temperature ranges where the Mott-insulating state develops anti-ferromagnetic correlations \cite{Mazurenko2017}, and the first observation of the internal structure of a polaron in this setting was reported recently \cite{Koepsell2019}. Per expectations, this experiment confirms that the carrier is dressed by local reduction, or even reversal, of spin-correlations. A comparison to the case of a pinned dopant establishes that delocalization is essential to the structure of the polaron.

Worm-algorithm Monte Carlo (WAMC) provides an exceptionally efficient protocol for obtaining the statistical properties of quantum many-body systems at thermal equilibrium \cite{Prokofev1998}. For bosonic systems, this technique allows computing unbiased and controllable results for strongly interacting systems at energy scales far below condensation \cite{PhysRevB.75.134302}. For fermions, the sign problem limits the applicability to Gutzwiller-projected theories where sign fluctuations are only extensive in the number of carriers as opposed to system size.
To date, this technique has only been used to derive the spectral properties of a single carrier in the $t$-$J$ model \cite{PhysRevB.18.3453}, which transpires at imaginary time-scales where the sign problem is irrelevant \cite{PhysRevB.64.033101}.
In this work, we employ WAMC to resolve the internal structure of a single polaron in the $t$-$J$ model. By relying on an efficient sampling protocol and large scale simulations, we can extract accurate data deep into the temperature range where the sign problem becomes significant. As a result, we obtain unbiased spin-correlations in the proximity of the charge carrier that are directly comparable to experiments on ultra-cold atomic gases. For a discussion of WAMC, see the supplementary information.

In our simulations, we represent fermions as world-lines in space and imaginary time. We apply periodic boundary conditions with a lattice size of $ 20 \times 20 $, which substantially exceeds the characteristic size of the polaron, ensuring that boundary pressure is not present.
Periodic boundary conditions are also implemented in the imaginary-time direction, reflecting that the trace only involves diagonal elements of the density matrix.
We use separate worms for the spin and charge sectors \cite{Prokofev1998}. The former can wind in imaginary time, which alters the total spin in the system so that this sector is in the grand canonical ensemble.
The worm corresponding to charge cannot wind, keeping the number of carriers to unity at all times. We confirm the accuracy of this technique by comparison to ED on a $ 4 \times 4 $ lattice, see supplementary information.

In this setup, the sign problem appears as fermions may swap positions through a combination of hopping and super-exchange processes in the background. The first process which gives a sign change is of the order $t^2 J^3$, and becomes apparent at temperatures of $t\sim 0.05$. Relying on WAMC, we are able to resolve temperatures down to $t\approx 0.011$ with virtually vanishing error bars.

\begin{figure}[!htb]
  \includegraphics[width=\linewidth]{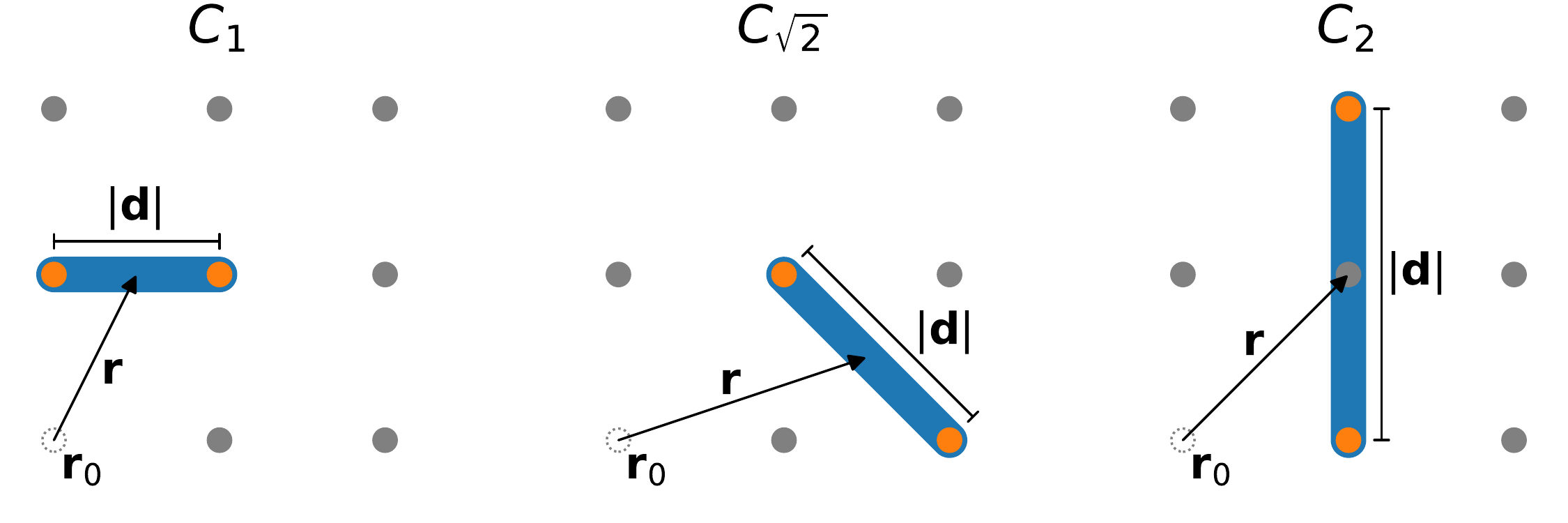}
  \caption{
    Illustration of the correlators $ C_1 $, $ C_{\sqrt 2} $, and $ C_2 $ defined in Eq. (\ref{correlators}). Here, the bond distance $\br$ is defined as the distance from the carrier to the point between the spins for which the correlation is considered.
  }
  \label{CX_illustrations}
\end{figure}

We focus on observing real-space spin-correlations, as these are also relevant for current experiments based on electron gas microscopy. In line with \cite{Koepsell2019}, we consider correlators of the form
\bea
\label{correlators}
C_{|\mbf d|}(\mbf r) = 4 \langle S^z_{\mbf r_0 + \mbf r - \mbf d / 2} \, S^z_{\mbf r_0 + \mbf r + \mbf d / 2} \rangle.
\eea
Here $ \mbf r_0 $ and $ \mbf r_0 + \mbf r \pm \mbf d / 2 $ are the positions of the carrier and the two spins, respectively, while $\br$ defines the bond distance. We will consider only the cases $ | \mbf d | = 1 $ (nearest-neighbor), $ | \mbf d | = \sqrt 2 $ (next-nearest-neighbor), and $ | \mbf d | = 2 $ (next-next-nearest-neighbor). The correlators are illustrated in Fig.\ \ref{CX_illustrations}.
The principal observables generated by WAMC are diagonal elements of the density matrix,
from which we can readily extract the correlators (\ref{correlators}).

We describe the polarons using the $t$-$J$ model, which captures the low-energy physics of the Hubbard model at large on-site repulsion, which is relevant for a doped Mott insulator.
We set the super-exchange to $ J/t=0.3 $, corresponding to an energy scale of the contact interaction given by $U/t=4t/J\approx 13.33$. This parameter choice is directly comparable to the experiment \cite{Koepsell2019}, and puts us well into the strongly correlated regime.

\begin{figure}[!htb]
  \includegraphics[width=\linewidth]{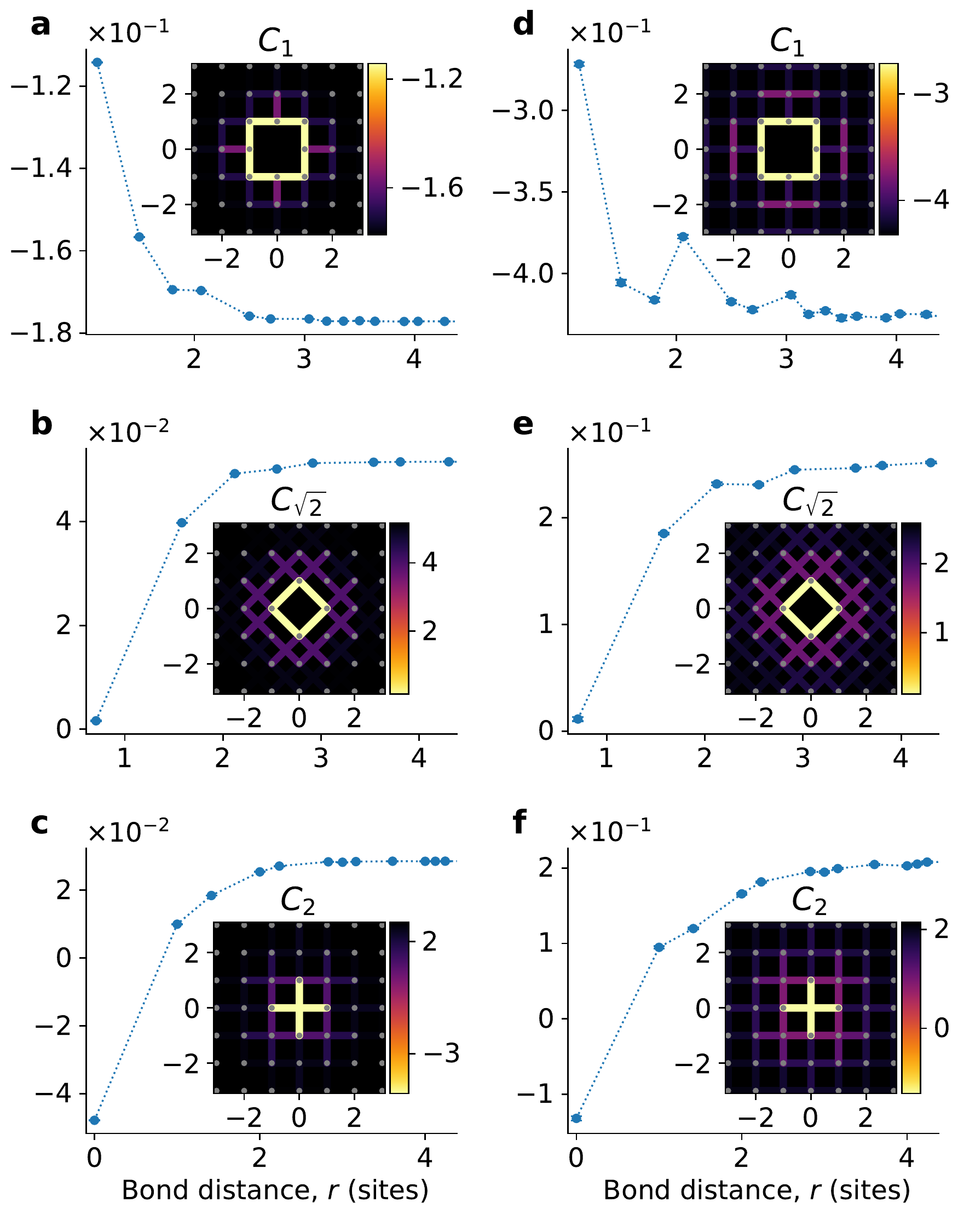}
  \caption{
  Spin-correlations as a function of the bond distance, obtained for a system with $20\times 20$ lattice sites and periodic boundary conditions. Here, $J/t=0.3$ which corresponds to $U/t\approx 13.33$, while temperatures are given by $T=t/2.2$ (a-c) and $T=t/8.8$ (d-f) respectively.
  The nearest-neighbor correlations (a,d) reveal a reduction of anti-ferromagnetism, though the lack of ferromagnetism implies no significant increase in the mean-free path. Next nearest-neighbor correlations (b,e) vanish near the carrier, implying a high degree of frustration that persists to low temperature. The combination of super-exchange and delocalization gives rise to anti-correlation across the carrier (c,f). While error bars are given for this data, they are not discernible to the naked eye. 
  }
  \label{panel}
\end{figure}

The spin-correlators (\ref{correlators}) obtained from WAMC are shown in Fig. \ref{panel}. Plots (a-c) correspond to a temperature of $T=t/2.2$, which lies within the error bars of the experiment \cite{Koepsell2019}.
Like that work, we find a reduction of $C_1$ (a) near the carrier, though substantial anti-ferromagnetism persists even in its direct vicinity.
The effect on $C_{\sqrt 2}$ (b) is more dramatic than on $C_1$, with correlations virtually disappearing at $r=\sqrt{1/2}$. By comparison, experimental findings even indicate anti-correlation close to the carrier.
For $C_2$, we find a dramatic reversal, with strong anti-correlation appearing across the carrier (c), in line with observations.
A comparison of theoretical and experimental results thus indicates that the $t$-$J$ model qualitatively captures the significant features of the polaron, with some quantitative discrepancies. We also find qualitative agreement with zero-temperature calculations based on DMRG and trial wave-function methods \cite{PhysRevB.99.224422}.

Reducing the temperature by a factor $4$ (d-f), the magnitude of the spin-correlations increases significantly, with $C_1$ approaching its ground state value on the background \cite{PhysRevA.84.053611}. The spatial extent of the polaron grows somewhat, while the basic features of its internal structure persist.

The reversal of  $C_2$  and the reduction of $C_{\sqrt 2}$ result from competing super-exchange processes that are activated as the carrier moves through the background. When the dopant hops a lattice spacing, spins that where previously nearest or next nearest-neighbors--and therefore correlated--are brought in direct contact where the interaction is antiferromagnetic. From (b,e), it is clear that this leads to a highly frustrated state with almost vanishing correlations that also persist at low temperatures.

\begin{figure}[!htb]
  \includegraphics[width=\linewidth]{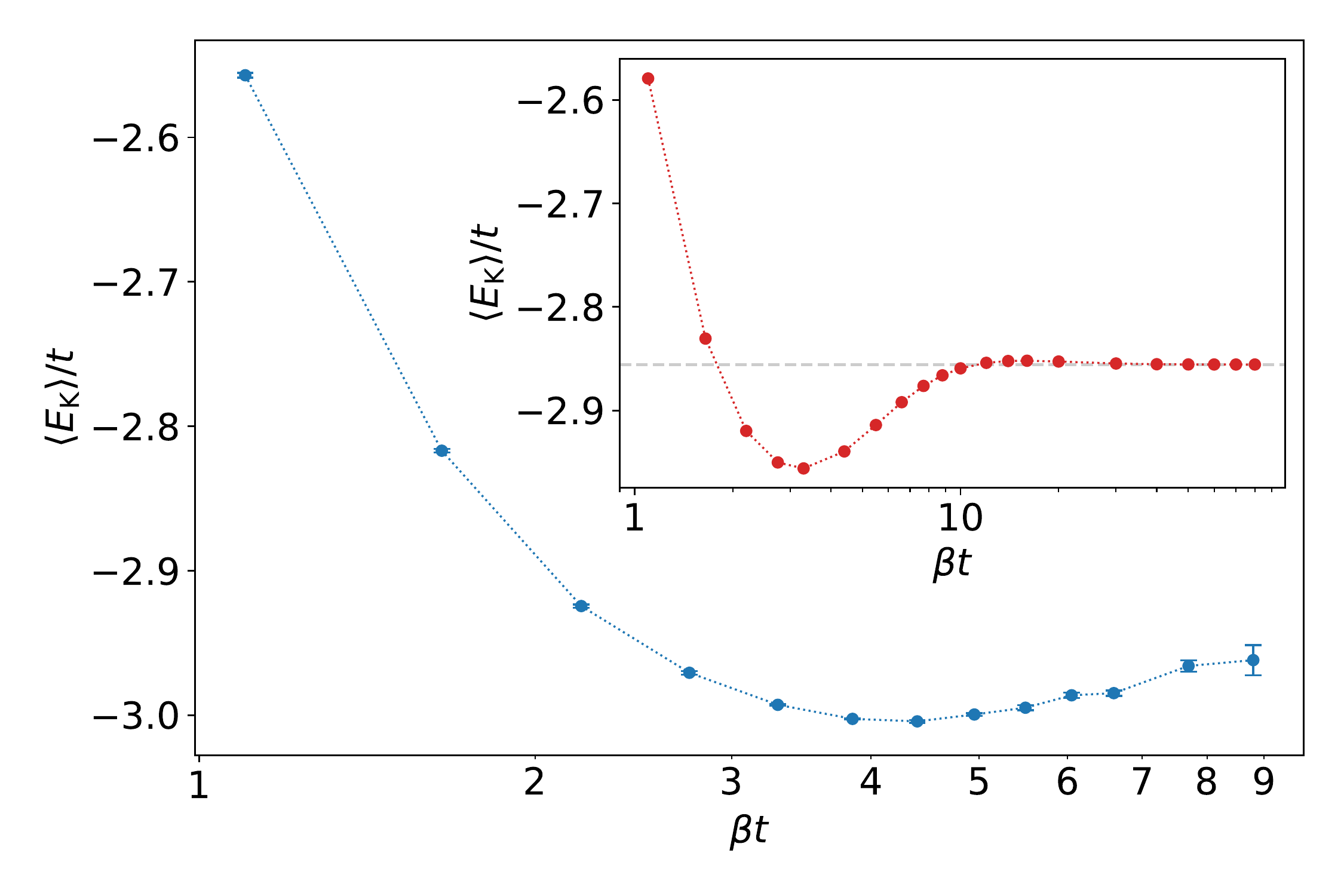}
  \caption{Kinetic energy of the polaron as a function of the inverse temperature. At lower temperatures, magnetic correlations build up in the background, increasing the energy cost of delocalization. This is reflected in the the non-monotonic dependence of kinetic energy on temperature. The inset shows kinetic energy for a $4\times 4$ system, obtained via ED. 
}
  \label{energy}
\end{figure}

The spin configuration of the polaron does not provide a significant increase of the mean-free path, which would require creating ferromagnetic correlations in the proximity of the carrier. This implies a substantial suppression of the density of states at the band ends \cite{brink}. The resulting correlations thus indicate competition between super-exchange and processes that also involve propagation of the carrier. Purely kinetic mechanisms that drive the system towards an increased mean-free path \cite{Andreev} are not visible in this data.
This picture is further corroborated by the kinetic energy of the polaron, which is shown in Fig. \ref{energy}. The frustration which is visible in Fig. \ref{panel} (b,e) becomes increasingly important at low temperatures where the system develops strong magnetic correlations. This makes delocalization of the carrier energetically expensive, eventually leading to increasing kinetic energy below $\beta t \approx 0.44$.

 We have presented unbiased high-precision results for the real-space structure of a polaron in the $t$-$J$ model. Our results reveal a distorted background that is characterized by reversal or depletion of spin-correlations near the carrier, which is indicative of competing magnetic processes and frustration. 
 We find a good agreement with recent quantum gas microscopy experiments, with minor discrepancies suggesting slight differences between our model and the physical system realized in \cite{Koepsell2019}. 
We have thus reached a point where direct comparison of single-particle measurements and essentially approximation-free theoretical techniques is possible. 
  With access to temperatures significantly below currently published experimental results, we can reach the regime where short-range correlations approach their ground state values. At the next stage, WAMC can be used to compute correlations between pairs of carriers, thereby providing critical insights into the temperature and energy scale of pairing. This can be carried out in tandem with experiments on multi-polaron structure formation in ultra-cold atomic gases, that have now become possible \cite{Chiu251}.

{\it Acknowledgments---}
This work was supported by the Swedish Research Council (VR). It was further supported by G{\aa}l\"ostiftelsen, through a travel stipend. Computations were performed on resources provided by the Swedish National Infrastructure for Computing (SNIC) at the National Supercomputer Centre in Link\"oping, Sweden.

\bibliography{biblio}

\begin{thebibliography}{7}%
\makeatletter
\providecommand \@ifxundefined [1]{%
 \@ifx{#1\undefined}
}%
\providecommand \@ifnum [1]{%
 \ifnum #1\expandafter \@firstoftwo
 \else \expandafter \@secondoftwo
 \fi
}%
\providecommand \@ifx [1]{%
 \ifx #1\expandafter \@firstoftwo
 \else \expandafter \@secondoftwo
 \fi
}%
\providecommand \natexlab [1]{#1}%
\providecommand \enquote  [1]{``#1''}%
\providecommand \bibnamefont  [1]{#1}%
\providecommand \bibfnamefont [1]{#1}%
\providecommand \citenamefont [1]{#1}%
\providecommand \href@noop [0]{\@secondoftwo}%
\providecommand \href [0]{\begingroup \@sanitize@url \@href}%
\providecommand \@href[1]{\@@startlink{#1}\@@href}%
\providecommand \@@href[1]{\endgroup#1\@@endlink}%
\providecommand \@sanitize@url [0]{\catcode `\\12\catcode `\$12\catcode
  `\&12\catcode `\#12\catcode `\^12\catcode `\_12\catcode `\%12\relax}%
\providecommand \@@startlink[1]{}%
\providecommand \@@endlink[0]{}%
\providecommand \url  [0]{\begingroup\@sanitize@url \@url }%
\providecommand \@url [1]{\endgroup\@href {#1}{\urlprefix }}%
\providecommand \urlprefix  [0]{URL }%
\providecommand \Eprint [0]{\href }%
\providecommand \doibase [0]{http://dx.doi.org/}%
\providecommand \selectlanguage [0]{\@gobble}%
\providecommand \bibinfo  [0]{\@secondoftwo}%
\providecommand \bibfield  [0]{\@secondoftwo}%
\providecommand \translation [1]{[#1]}%
\providecommand \BibitemOpen [0]{}%
\providecommand \bibitemStop [0]{}%
\providecommand \bibitemNoStop [0]{.\EOS\space}%
\providecommand \EOS [0]{\spacefactor3000\relax}%
\providecommand \BibitemShut  [1]{\csname bibitem#1\endcsname}%
\let\auto@bib@innerbib\@empty
\bibitem [{\citenamefont {Spałek}(2007)}]{spalek}%
  \BibitemOpen
  \bibfield  {author} {\bibinfo {author} {\bibfnamefont {J.}~\bibnamefont
  {Spałek}},\ }\href {\doibase 10.12693/APhysPolA.111.409} {\bibfield
  {journal} {\bibinfo  {journal} {Acta Physica Polonica A}\ }\textbf {\bibinfo
  {volume} {111}},\ \bibinfo {pages} {409} (\bibinfo {year}
  {2007})}\BibitemShut {NoStop}%
\bibitem [{\citenamefont {Hubbard}(1963)}]{hubbard}%
  \BibitemOpen
  \bibfield  {author} {\bibinfo {author} {\bibfnamefont {J.}~\bibnamefont
  {Hubbard}},\ }\href {http://www.jstor.org/stable/2414761} {\bibfield
  {journal} {\bibinfo  {journal} {Proceedings of the Royal Society of London.
  Series A, Mathematical and Physical Sciences}\ }\textbf {\bibinfo {volume}
  {276}},\ \bibinfo {pages} {238} (\bibinfo {year} {1963})}\BibitemShut
  {NoStop}%
\bibitem [{\citenamefont {Prokof'ev}\ \emph {et~al.}(1998)\citenamefont
  {Prokof'ev}, \citenamefont {Svistunov},\ and\ \citenamefont
  {Tupitsyn}}]{Prokofev1998}%
  \BibitemOpen
  \bibfield  {author} {\bibinfo {author} {\bibfnamefont {N.~V.}\ \bibnamefont
  {Prokof'ev}}, \bibinfo {author} {\bibfnamefont {B.~V.}\ \bibnamefont
  {Svistunov}}, \ and\ \bibinfo {author} {\bibfnamefont {I.~S.}\ \bibnamefont
  {Tupitsyn}},\ }\href {\doibase 10.1134/1.558661} {\bibfield  {journal}
  {\bibinfo  {journal} {Journal of Experimental and Theoretical Physics}\
  }\textbf {\bibinfo {volume} {87}},\ \bibinfo {pages} {310} (\bibinfo {year}
  {1998})}\BibitemShut {NoStop}%
\bibitem [{\citenamefont {Sellin}(2018)}]{Sellin1173323}%
  \BibitemOpen
  \bibfield  {author} {\bibinfo {author} {\bibfnamefont {K.}~\bibnamefont
  {Sellin}},\ }\emph {\bibinfo {title} {Structure formation, phase transitions
  and drag interactions in multicomponent superconductors and superfluids}},\
  \href {http://www.diva-portal.org/smash/get/diva2:1173323/FULLTEXT01.pdf}
  {Ph.D. thesis},\ \bibinfo  {school} {KTH, Statistical Physics} (\bibinfo
  {year} {2018})\BibitemShut {NoStop}%
\bibitem [{\citenamefont {Troyer}\ and\ \citenamefont
  {Wiese}(2005)}]{PhysRevLett.94.170201}%
  \BibitemOpen
  \bibfield  {author} {\bibinfo {author} {\bibfnamefont {M.}~\bibnamefont
  {Troyer}}\ and\ \bibinfo {author} {\bibfnamefont {U.-J.}\ \bibnamefont
  {Wiese}},\ }\href {\doibase 10.1103/PhysRevLett.94.170201} {\bibfield
  {journal} {\bibinfo  {journal} {Phys. Rev. Lett.}\ }\textbf {\bibinfo
  {volume} {94}},\ \bibinfo {pages} {170201} (\bibinfo {year}
  {2005})}\BibitemShut {NoStop}%
\bibitem [{\citenamefont {Mishchenko}\ \emph {et~al.}(2001)\citenamefont
  {Mishchenko}, \citenamefont {Prokof'ev},\ and\ \citenamefont
  {Svistunov}}]{PhysRevB.64.033101}%
  \BibitemOpen
  \bibfield  {author} {\bibinfo {author} {\bibfnamefont {A.~S.}\ \bibnamefont
  {Mishchenko}}, \bibinfo {author} {\bibfnamefont {N.~V.}\ \bibnamefont
  {Prokof'ev}}, \ and\ \bibinfo {author} {\bibfnamefont {B.~V.}\ \bibnamefont
  {Svistunov}},\ }\href {\doibase 10.1103/PhysRevB.64.033101} {\bibfield
  {journal} {\bibinfo  {journal} {Phys. Rev. B}\ }\textbf {\bibinfo {volume}
  {64}},\ \bibinfo {pages} {033101} (\bibinfo {year} {2001})}\BibitemShut
  {NoStop}%
\bibitem [{\citenamefont {Sandvik}(2010)}]{doi:10.1063/1.3518900}%
  \BibitemOpen
  \bibfield  {author} {\bibinfo {author} {\bibfnamefont {A.~W.}\ \bibnamefont
  {Sandvik}},\ }\href {\doibase 10.1063/1.3518900} {\bibfield  {journal}
  {\bibinfo  {journal} {AIP Conference Proceedings}\ }\textbf {\bibinfo
  {volume} {1297}},\ \bibinfo {pages} {135} (\bibinfo {year} {2010})},\ \Eprint
  {http://arxiv.org/abs/https://aip.scitation.org/doi/pdf/10.1063/1.3518900}
  {https://aip.scitation.org/doi/pdf/10.1063/1.3518900} \BibitemShut {NoStop}%
\end{thebibliography}%


\begin{thebibliography}{23}%
\makeatletter
\providecommand \@ifxundefined [1]{%
 \@ifx{#1\undefined}
}%
\providecommand \@ifnum [1]{%
 \ifnum #1\expandafter \@firstoftwo
 \else \expandafter \@secondoftwo
 \fi
}%
\providecommand \@ifx [1]{%
 \ifx #1\expandafter \@firstoftwo
 \else \expandafter \@secondoftwo
 \fi
}%
\providecommand \natexlab [1]{#1}%
\providecommand \enquote  [1]{``#1''}%
\providecommand \bibnamefont  [1]{#1}%
\providecommand \bibfnamefont [1]{#1}%
\providecommand \citenamefont [1]{#1}%
\providecommand \href@noop [0]{\@secondoftwo}%
\providecommand \href [0]{\begingroup \@sanitize@url \@href}%
\providecommand \@href[1]{\@@startlink{#1}\@@href}%
\providecommand \@@href[1]{\endgroup#1\@@endlink}%
\providecommand \@sanitize@url [0]{\catcode `\\12\catcode `\$12\catcode
  `\&12\catcode `\#12\catcode `\^12\catcode `\_12\catcode `\%12\relax}%
\providecommand \@@startlink[1]{}%
\providecommand \@@endlink[0]{}%
\providecommand \url  [0]{\begingroup\@sanitize@url \@url }%
\providecommand \@url [1]{\endgroup\@href {#1}{\urlprefix }}%
\providecommand \urlprefix  [0]{URL }%
\providecommand \Eprint [0]{\href }%
\providecommand \doibase [0]{http://dx.doi.org/}%
\providecommand \selectlanguage [0]{\@gobble}%
\providecommand \bibinfo  [0]{\@secondoftwo}%
\providecommand \bibfield  [0]{\@secondoftwo}%
\providecommand \translation [1]{[#1]}%
\providecommand \BibitemOpen [0]{}%
\providecommand \bibitemStop [0]{}%
\providecommand \bibitemNoStop [0]{.\EOS\space}%
\providecommand \EOS [0]{\spacefactor3000\relax}%
\providecommand \BibitemShut  [1]{\csname bibitem#1\endcsname}%
\let\auto@bib@innerbib\@empty
\bibitem [{\citenamefont {Alexandrov}\ and\ \citenamefont
  {Devreese}(2010)}]{Polaron}%
  \BibitemOpen
  \bibfield  {author} {\bibinfo {author} {\bibfnamefont {S.}~\bibnamefont
  {Alexandrov}}\ and\ \bibinfo {author} {\bibfnamefont {J.~T.}\ \bibnamefont
  {Devreese}},\ }\href@noop {} {\emph {\bibinfo {title} {Advances in Polaron
  Physics}}}\ (\bibinfo  {publisher} {Springer-Verlag Berlin Heidelberg},\
  \bibinfo {year} {2010})\BibitemShut {NoStop}%
\bibitem [{\citenamefont {Kane}\ \emph {et~al.}(1989)\citenamefont {Kane},
  \citenamefont {Lee},\ and\ \citenamefont {Read}}]{PhysRevB.39.6880}%
  \BibitemOpen
  \bibfield  {author} {\bibinfo {author} {\bibfnamefont {C.~L.}\ \bibnamefont
  {Kane}}, \bibinfo {author} {\bibfnamefont {P.~A.}\ \bibnamefont {Lee}}, \
  and\ \bibinfo {author} {\bibfnamefont {N.}~\bibnamefont {Read}},\ }\href
  {\doibase 10.1103/PhysRevB.39.6880} {\bibfield  {journal} {\bibinfo
  {journal} {Phys. Rev. B}\ }\textbf {\bibinfo {volume} {39}},\ \bibinfo
  {pages} {6880} (\bibinfo {year} {1989})}\BibitemShut {NoStop}%
\bibitem [{\citenamefont {Carlstr\"om}\ \emph {et~al.}(2016)\citenamefont
  {Carlstr\"om}, \citenamefont {Prokof'ev},\ and\ \citenamefont
  {Svistunov}}]{PhysRevLett.116.247202}%
  \BibitemOpen
  \bibfield  {author} {\bibinfo {author} {\bibfnamefont {J.}~\bibnamefont
  {Carlstr\"om}}, \bibinfo {author} {\bibfnamefont {N.}~\bibnamefont
  {Prokof'ev}}, \ and\ \bibinfo {author} {\bibfnamefont {B.}~\bibnamefont
  {Svistunov}},\ }\href {\doibase 10.1103/PhysRevLett.116.247202} {\bibfield
  {journal} {\bibinfo  {journal} {Phys. Rev. Lett.}\ }\textbf {\bibinfo
  {volume} {116}},\ \bibinfo {pages} {247202} (\bibinfo {year}
  {2016})}\BibitemShut {NoStop}%
\bibitem [{\citenamefont {Trugman}(1988)}]{PhysRevB.37.1597}%
  \BibitemOpen
  \bibfield  {author} {\bibinfo {author} {\bibfnamefont {S.~A.}\ \bibnamefont
  {Trugman}},\ }\href {\doibase 10.1103/PhysRevB.37.1597} {\bibfield  {journal}
  {\bibinfo  {journal} {Phys. Rev. B}\ }\textbf {\bibinfo {volume} {37}},\
  \bibinfo {pages} {1597} (\bibinfo {year} {1988})}\BibitemShut {NoStop}%
\bibitem [{\citenamefont {Koepsell}\ \emph {et~al.}(2019)\citenamefont
  {Koepsell}, \citenamefont {Vijayan}, \citenamefont {Sompet}, \citenamefont
  {Grusdt}, \citenamefont {Hilker}, \citenamefont {Demler}, \citenamefont
  {Salomon}, \citenamefont {Bloch},\ and\ \citenamefont
  {Gross}}]{Koepsell2019}%
  \BibitemOpen
  \bibfield  {author} {\bibinfo {author} {\bibfnamefont {J.}~\bibnamefont
  {Koepsell}}, \bibinfo {author} {\bibfnamefont {J.}~\bibnamefont {Vijayan}},
  \bibinfo {author} {\bibfnamefont {P.}~\bibnamefont {Sompet}}, \bibinfo
  {author} {\bibfnamefont {F.}~\bibnamefont {Grusdt}}, \bibinfo {author}
  {\bibfnamefont {T.~A.}\ \bibnamefont {Hilker}}, \bibinfo {author}
  {\bibfnamefont {E.}~\bibnamefont {Demler}}, \bibinfo {author} {\bibfnamefont
  {G.}~\bibnamefont {Salomon}}, \bibinfo {author} {\bibfnamefont
  {I.}~\bibnamefont {Bloch}}, \ and\ \bibinfo {author} {\bibfnamefont
  {C.}~\bibnamefont {Gross}},\ }\href {\doibase 10.1038/s41586-019-1463-1}
  {\bibfield  {journal} {\bibinfo  {journal} {Nature}\ }\textbf {\bibinfo
  {volume} {572}},\ \bibinfo {pages} {358} (\bibinfo {year}
  {2019})}\BibitemShut {NoStop}%
\bibitem [{\citenamefont {Lee}\ \emph {et~al.}(2006)\citenamefont {Lee},
  \citenamefont {Nagaosa},\ and\ \citenamefont {Wen}}]{RevModPhys.78.17}%
  \BibitemOpen
  \bibfield  {author} {\bibinfo {author} {\bibfnamefont {P.~A.}\ \bibnamefont
  {Lee}}, \bibinfo {author} {\bibfnamefont {N.}~\bibnamefont {Nagaosa}}, \ and\
  \bibinfo {author} {\bibfnamefont {X.-G.}\ \bibnamefont {Wen}},\ }\href
  {\doibase 10.1103/RevModPhys.78.17} {\bibfield  {journal} {\bibinfo
  {journal} {Rev. Mod. Phys.}\ }\textbf {\bibinfo {volume} {78}},\ \bibinfo
  {pages} {17} (\bibinfo {year} {2006})}\BibitemShut {NoStop}%
\bibitem [{\citenamefont {de~Gennes}(1960)}]{PhysRev.118.141}%
  \BibitemOpen
  \bibfield  {author} {\bibinfo {author} {\bibfnamefont {P.~G.}\ \bibnamefont
  {de~Gennes}},\ }\href {\doibase 10.1103/PhysRev.118.141} {\bibfield
  {journal} {\bibinfo  {journal} {Phys. Rev.}\ }\textbf {\bibinfo {volume}
  {118}},\ \bibinfo {pages} {141} (\bibinfo {year} {1960})}\BibitemShut
  {NoStop}%
\bibitem [{\citenamefont {Brinkman}\ and\ \citenamefont {Rice}(1970)}]{brink}%
  \BibitemOpen
  \bibfield  {author} {\bibinfo {author} {\bibfnamefont {W.~F.}\ \bibnamefont
  {Brinkman}}\ and\ \bibinfo {author} {\bibfnamefont {T.~M.}\ \bibnamefont
  {Rice}},\ }\href {\doibase 10.1103/PhysRevB.2.1324} {\bibfield  {journal}
  {\bibinfo  {journal} {Phys. Rev. B}\ }\textbf {\bibinfo {volume} {2}},\
  \bibinfo {pages} {1324} (\bibinfo {year} {1970})}\BibitemShut {NoStop}%
\bibitem [{\citenamefont {ANDERSON}(1987)}]{ANDERSON1196}%
  \BibitemOpen
  \bibfield  {author} {\bibinfo {author} {\bibfnamefont {P.~W.}\ \bibnamefont
  {ANDERSON}},\ }\href {\doibase 10.1126/science.235.4793.1196} {\bibfield
  {journal} {\bibinfo  {journal} {Science}\ }\textbf {\bibinfo {volume}
  {235}},\ \bibinfo {pages} {1196} (\bibinfo {year} {1987})}\BibitemShut
  {NoStop}%
\bibitem [{\citenamefont {Schrieffer}\ \emph {et~al.}(1989)\citenamefont
  {Schrieffer}, \citenamefont {Wen},\ and\ \citenamefont
  {Zhang}}]{PhysRevB.39.11663}%
  \BibitemOpen
  \bibfield  {author} {\bibinfo {author} {\bibfnamefont {J.~R.}\ \bibnamefont
  {Schrieffer}}, \bibinfo {author} {\bibfnamefont {X.~G.}\ \bibnamefont {Wen}},
  \ and\ \bibinfo {author} {\bibfnamefont {S.~C.}\ \bibnamefont {Zhang}},\
  }\href {\doibase 10.1103/PhysRevB.39.11663} {\bibfield  {journal} {\bibinfo
  {journal} {Phys. Rev. B}\ }\textbf {\bibinfo {volume} {39}},\ \bibinfo
  {pages} {11663} (\bibinfo {year} {1989})}\BibitemShut {NoStop}%
\bibitem [{\citenamefont {White}\ and\ \citenamefont
  {Scalapino}(1997)}]{PhysRevB.55.6504}%
  \BibitemOpen
  \bibfield  {author} {\bibinfo {author} {\bibfnamefont {S.~R.}\ \bibnamefont
  {White}}\ and\ \bibinfo {author} {\bibfnamefont {D.~J.}\ \bibnamefont
  {Scalapino}},\ }\href {\doibase 10.1103/PhysRevB.55.6504} {\bibfield
  {journal} {\bibinfo  {journal} {Phys. Rev. B}\ }\textbf {\bibinfo {volume}
  {55}},\ \bibinfo {pages} {6504} (\bibinfo {year} {1997})}\BibitemShut
  {NoStop}%
\bibitem [{\citenamefont {White}(1993)}]{PhysRevB.48.10345}%
  \BibitemOpen
  \bibfield  {author} {\bibinfo {author} {\bibfnamefont {S.~R.}\ \bibnamefont
  {White}},\ }\href {\doibase 10.1103/PhysRevB.48.10345} {\bibfield  {journal}
  {\bibinfo  {journal} {Phys. Rev. B}\ }\textbf {\bibinfo {volume} {48}},\
  \bibinfo {pages} {10345} (\bibinfo {year} {1993})}\BibitemShut {NoStop}%
\bibitem [{\citenamefont {Loh}\ \emph {et~al.}(1990)\citenamefont {Loh},
  \citenamefont {Gubernatis}, \citenamefont {Scalettar}, \citenamefont {White},
  \citenamefont {Scalapino},\ and\ \citenamefont {Sugar}}]{PhysRevB.41.9301}%
  \BibitemOpen
  \bibfield  {author} {\bibinfo {author} {\bibfnamefont {E.~Y.}\ \bibnamefont
  {Loh}}, \bibinfo {author} {\bibfnamefont {J.~E.}\ \bibnamefont {Gubernatis}},
  \bibinfo {author} {\bibfnamefont {R.~T.}\ \bibnamefont {Scalettar}}, \bibinfo
  {author} {\bibfnamefont {S.~R.}\ \bibnamefont {White}}, \bibinfo {author}
  {\bibfnamefont {D.~J.}\ \bibnamefont {Scalapino}}, \ and\ \bibinfo {author}
  {\bibfnamefont {R.~L.}\ \bibnamefont {Sugar}},\ }\href {\doibase
  10.1103/PhysRevB.41.9301} {\bibfield  {journal} {\bibinfo  {journal} {Phys.
  Rev. B}\ }\textbf {\bibinfo {volume} {41}},\ \bibinfo {pages} {9301}
  (\bibinfo {year} {1990})}\BibitemShut {NoStop}%
\bibitem [{\citenamefont {Gross}\ and\ \citenamefont {Bloch}(2017)}]{Gross995}%
  \BibitemOpen
  \bibfield  {author} {\bibinfo {author} {\bibfnamefont {C.}~\bibnamefont
  {Gross}}\ and\ \bibinfo {author} {\bibfnamefont {I.}~\bibnamefont {Bloch}},\
  }\href {\doibase 10.1126/science.aal3837} {\bibfield  {journal} {\bibinfo
  {journal} {Science}\ }\textbf {\bibinfo {volume} {357}},\ \bibinfo {pages}
  {995} (\bibinfo {year} {2017})}\BibitemShut {NoStop}%
\bibitem [{\citenamefont {Mazurenko}\ \emph {et~al.}(2017)\citenamefont
  {Mazurenko}, \citenamefont {Chiu}, \citenamefont {Ji}, \citenamefont
  {Parsons}, \citenamefont {Kan{\'a}sz-Nagy}, \citenamefont {Schmidt},
  \citenamefont {Grusdt}, \citenamefont {Demler}, \citenamefont {Greif},\ and\
  \citenamefont {Greiner}}]{Mazurenko2017}%
  \BibitemOpen
  \bibfield  {author} {\bibinfo {author} {\bibfnamefont {A.}~\bibnamefont
  {Mazurenko}}, \bibinfo {author} {\bibfnamefont {C.~S.}\ \bibnamefont {Chiu}},
  \bibinfo {author} {\bibfnamefont {G.}~\bibnamefont {Ji}}, \bibinfo {author}
  {\bibfnamefont {M.~F.}\ \bibnamefont {Parsons}}, \bibinfo {author}
  {\bibfnamefont {M.}~\bibnamefont {Kan{\'a}sz-Nagy}}, \bibinfo {author}
  {\bibfnamefont {R.}~\bibnamefont {Schmidt}}, \bibinfo {author} {\bibfnamefont
  {F.}~\bibnamefont {Grusdt}}, \bibinfo {author} {\bibfnamefont
  {E.}~\bibnamefont {Demler}}, \bibinfo {author} {\bibfnamefont
  {D.}~\bibnamefont {Greif}}, \ and\ \bibinfo {author} {\bibfnamefont
  {M.}~\bibnamefont {Greiner}},\ }\href {\doibase 10.1038/nature22362}
  {\bibfield  {journal} {\bibinfo  {journal} {Nature}\ }\textbf {\bibinfo
  {volume} {545}},\ \bibinfo {pages} {462} (\bibinfo {year}
  {2017})}\BibitemShut {NoStop}%
\bibitem [{\citenamefont {Prokof'ev}\ \emph {et~al.}(1998)\citenamefont
  {Prokof'ev}, \citenamefont {Svistunov},\ and\ \citenamefont
  {Tupitsyn}}]{Prokofev1998}%
  \BibitemOpen
  \bibfield  {author} {\bibinfo {author} {\bibfnamefont {N.~V.}\ \bibnamefont
  {Prokof'ev}}, \bibinfo {author} {\bibfnamefont {B.~V.}\ \bibnamefont
  {Svistunov}}, \ and\ \bibinfo {author} {\bibfnamefont {I.~S.}\ \bibnamefont
  {Tupitsyn}},\ }\href {\doibase 10.1134/1.558661} {\bibfield  {journal}
  {\bibinfo  {journal} {Journal of Experimental and Theoretical Physics}\
  }\textbf {\bibinfo {volume} {87}},\ \bibinfo {pages} {310} (\bibinfo {year}
  {1998})}\BibitemShut {NoStop}%
\bibitem [{\citenamefont {Capogrosso-Sansone}\ \emph
  {et~al.}(2007)\citenamefont {Capogrosso-Sansone}, \citenamefont {Prokof'ev},\
  and\ \citenamefont {Svistunov}}]{PhysRevB.75.134302}%
  \BibitemOpen
  \bibfield  {author} {\bibinfo {author} {\bibfnamefont {B.}~\bibnamefont
  {Capogrosso-Sansone}}, \bibinfo {author} {\bibfnamefont {N.~V.}\ \bibnamefont
  {Prokof'ev}}, \ and\ \bibinfo {author} {\bibfnamefont {B.~V.}\ \bibnamefont
  {Svistunov}},\ }\href {\doibase 10.1103/PhysRevB.75.134302} {\bibfield
  {journal} {\bibinfo  {journal} {Phys. Rev. B}\ }\textbf {\bibinfo {volume}
  {75}},\ \bibinfo {pages} {134302} (\bibinfo {year} {2007})}\BibitemShut
  {NoStop}%
\bibitem [{\citenamefont {Chao}\ \emph {et~al.}(1978)\citenamefont {Chao},
  \citenamefont {Spa\l{}ek},\ and\ \citenamefont {Ole\ifmmode~\acute{s}\else
  \'{s}\fi{}}}]{PhysRevB.18.3453}%
  \BibitemOpen
  \bibfield  {author} {\bibinfo {author} {\bibfnamefont {K.~A.}\ \bibnamefont
  {Chao}}, \bibinfo {author} {\bibfnamefont {J.}~\bibnamefont {Spa\l{}ek}}, \
  and\ \bibinfo {author} {\bibfnamefont {A.~M.}\ \bibnamefont
  {Ole\ifmmode~\acute{s}\else \'{s}\fi{}}},\ }\href {\doibase
  10.1103/PhysRevB.18.3453} {\bibfield  {journal} {\bibinfo  {journal} {Phys.
  Rev. B}\ }\textbf {\bibinfo {volume} {18}},\ \bibinfo {pages} {3453}
  (\bibinfo {year} {1978})}\BibitemShut {NoStop}%
\bibitem [{\citenamefont {Mishchenko}\ \emph {et~al.}(2001)\citenamefont
  {Mishchenko}, \citenamefont {Prokof'ev},\ and\ \citenamefont
  {Svistunov}}]{PhysRevB.64.033101}%
  \BibitemOpen
  \bibfield  {author} {\bibinfo {author} {\bibfnamefont {A.~S.}\ \bibnamefont
  {Mishchenko}}, \bibinfo {author} {\bibfnamefont {N.~V.}\ \bibnamefont
  {Prokof'ev}}, \ and\ \bibinfo {author} {\bibfnamefont {B.~V.}\ \bibnamefont
  {Svistunov}},\ }\href {\doibase 10.1103/PhysRevB.64.033101} {\bibfield
  {journal} {\bibinfo  {journal} {Phys. Rev. B}\ }\textbf {\bibinfo {volume}
  {64}},\ \bibinfo {pages} {033101} (\bibinfo {year} {2001})}\BibitemShut
  {NoStop}%
\bibitem [{\citenamefont {Grusdt}\ \emph {et~al.}(2019)\citenamefont {Grusdt},
  \citenamefont {Bohrdt},\ and\ \citenamefont {Demler}}]{PhysRevB.99.224422}%
  \BibitemOpen
  \bibfield  {author} {\bibinfo {author} {\bibfnamefont {F.}~\bibnamefont
  {Grusdt}}, \bibinfo {author} {\bibfnamefont {A.}~\bibnamefont {Bohrdt}}, \
  and\ \bibinfo {author} {\bibfnamefont {E.}~\bibnamefont {Demler}},\ }\href
  {\doibase 10.1103/PhysRevB.99.224422} {\bibfield  {journal} {\bibinfo
  {journal} {Phys. Rev. B}\ }\textbf {\bibinfo {volume} {99}},\ \bibinfo
  {pages} {224422} (\bibinfo {year} {2019})}\BibitemShut {NoStop}%
\bibitem [{\citenamefont {Khatami}\ and\ \citenamefont
  {Rigol}(2011)}]{PhysRevA.84.053611}%
  \BibitemOpen
  \bibfield  {author} {\bibinfo {author} {\bibfnamefont {E.}~\bibnamefont
  {Khatami}}\ and\ \bibinfo {author} {\bibfnamefont {M.}~\bibnamefont
  {Rigol}},\ }\href {\doibase 10.1103/PhysRevA.84.053611} {\bibfield  {journal}
  {\bibinfo  {journal} {Phys. Rev. A}\ }\textbf {\bibinfo {volume} {84}},\
  \bibinfo {pages} {053611} (\bibinfo {year} {2011})}\BibitemShut {NoStop}%
\bibitem [{\citenamefont {{Andreev}}(1976)}]{Andreev}%
  \BibitemOpen
  \bibfield  {author} {\bibinfo {author} {\bibfnamefont {A.~F.}\ \bibnamefont
  {{Andreev}}},\ }\href@noop {} {\bibfield  {journal} {\bibinfo  {journal}
  {Soviet Journal of Experimental and Theoretical Physics Letters}\ }\textbf
  {\bibinfo {volume} {24}},\ \bibinfo {pages} {564} (\bibinfo {year}
  {1976})}\BibitemShut {NoStop}%
\bibitem [{\citenamefont {Chiu}\ \emph {et~al.}(2019)\citenamefont {Chiu},
  \citenamefont {Ji}, \citenamefont {Bohrdt}, \citenamefont {Xu}, \citenamefont
  {Knap}, \citenamefont {Demler}, \citenamefont {Grusdt}, \citenamefont
  {Greiner},\ and\ \citenamefont {Greif}}]{Chiu251}%
  \BibitemOpen
  \bibfield  {author} {\bibinfo {author} {\bibfnamefont {C.~S.}\ \bibnamefont
  {Chiu}}, \bibinfo {author} {\bibfnamefont {G.}~\bibnamefont {Ji}}, \bibinfo
  {author} {\bibfnamefont {A.}~\bibnamefont {Bohrdt}}, \bibinfo {author}
  {\bibfnamefont {M.}~\bibnamefont {Xu}}, \bibinfo {author} {\bibfnamefont
  {M.}~\bibnamefont {Knap}}, \bibinfo {author} {\bibfnamefont {E.}~\bibnamefont
  {Demler}}, \bibinfo {author} {\bibfnamefont {F.}~\bibnamefont {Grusdt}},
  \bibinfo {author} {\bibfnamefont {M.}~\bibnamefont {Greiner}}, \ and\
  \bibinfo {author} {\bibfnamefont {D.}~\bibnamefont {Greif}},\ }\href
  {\doibase 10.1126/science.aav3587} {\bibfield  {journal} {\bibinfo  {journal}
  {Science}\ }\textbf {\bibinfo {volume} {365}},\ \bibinfo {pages} {251}
  (\bibinfo {year} {2019})}\BibitemShut {NoStop}%
\end{thebibliography}%

\end{document}


\title{Supplementary Information}

\author{Emil Blomquist$^{1}$ and Johan~Carlstr\"om$^{2}$ }
\affiliation{$^{1}$Department of Physics, Royal Institute of Technology, Stockholm, SE-106 91, Sweden}
\affiliation{$^{2}$Department of Physics, Stockholm University, 106 91 Stockholm, Sweden}
\date{\today}

\maketitle

\textbf{Model.}
We describe our system using the $t$-$J$ model \cite{spalek}, which captures the low energy physics of  the Hubbard model \cite{hubbard} in the $ U \gg t $ limit. We consider the case of a single carrier on a Mott-background. The Hamiltonian is then given by
\begin{equation}
  \begin{split}
    \hat H
    &=
    - t \sum _{\langle ij \rangle, \,\sigma}
    \hat a_{i \sigma}^\dagger (1 - \hat n_{i \bar \sigma})
    \hat a_{j \sigma} (1 - \hat n_{j \bar \sigma}) \\
    &+ J \sum _{\langle ij \rangle}
    \left(
    \hat{\mbf S}_i \cdot \hat{\mbf S}_j
    -
    \frac{\hat{n}_i\hat{n}_j}{4}
    \right)
  \end{split}
\end{equation}
%
%
where doubly occupied sites are forbidden. Here the parameter $ J $ is related to the Hubbard on-site interaction strength through $ J = 4 t^2 / U $.

\vspace{0.5em}
\textbf{The worm algorithm.}
The correlators and kinetic energy discussed in the main text were obtained using worm-algorithm Monte Carlo (WAMC) in a continuous-time setting \cite{Prokofev1998}. In this method, the partition- and Green's function sectors are combined into a single configuration space. Then, in order to sample partition function configurations, one is required to pass through the Green's function sector. Performing the simulations in real-space and imaginary-time, both the kinetic energy \cite{Sellin1173323} and the correlators of interest are directly accessible from the partition function.

\vspace{0.5em}
\textbf{Sign problem.}
Treating lattice fermions using WAMC, the sign problem \cite{PhysRevLett.94.170201} is unavoidable, despite simulating merely a single carrier. In order to account for the possibility of having negative configuration weights $ p = s |p| $, the sign $ s $ is kept track of while simulating the bosonic system, using as weight the aforementioned $ |p| $. The fermionic expectation value $ \langle A \rangle_\text{F} $ is then obtained from bosonic ones according to $ \langle A \rangle_\text{F} = \langle s A \rangle_\text{B} / \langle s \rangle_\text{B} $.

A process able to exchange a pair of indistinguishable fermions, and thus flip the sign, is present first at order $ t^2 J^3 $ \cite{PhysRevB.64.033101}. A world-line configuration containing processes of this type is rare at high temperatures but becomes more frequent when the temperature is lowered.
The typical signature of this is the exponential decay of the average sign $ \langle s \rangle_\text{B} $ as the temperature is decreased. Consequently, the signal-to-noise ratio will decrease exponentially with decreased temperature, which is the hallmark of the sign problem.

\begin{figure}[!htb]
  \includegraphics[width=1.\linewidth]{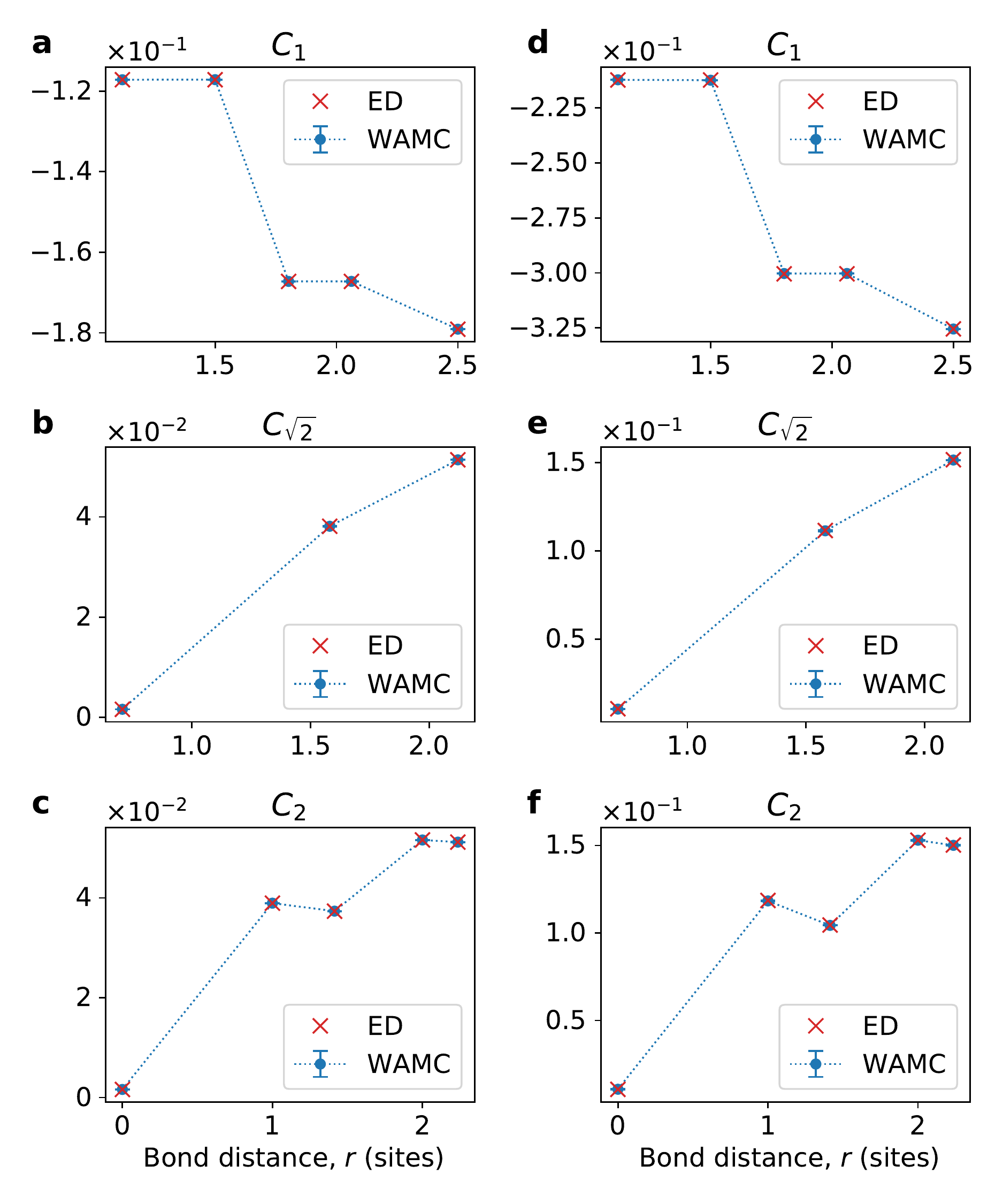}
  \caption{
    Spin-correlations $ C_1 $ (a,d), $ C_{ \sqrt 2 } $ (b,e), and $ C_2 $ (c,f) obtained via WAMC and ED, as a function of the distance from the carrier obtained for a system with $ 4 \times 4 $ lattice sites and periodic boundary conditions. Here $ J/t = 0.3 $ while temperatures are given by $ T = t/2.2 $ (a-c) and $ T = t/4.4 $ (d-f), respectively. These plots indicate a perfect agreement between the results obtained from ED and WAMC.
    While error bars are given for the WAMC data, they are smaller than the marker size.
  }
  \label{fig:CX_exact_vs_MC}
\end{figure}

\begin{figure}[!htb]
  \includegraphics[width=1.\linewidth]{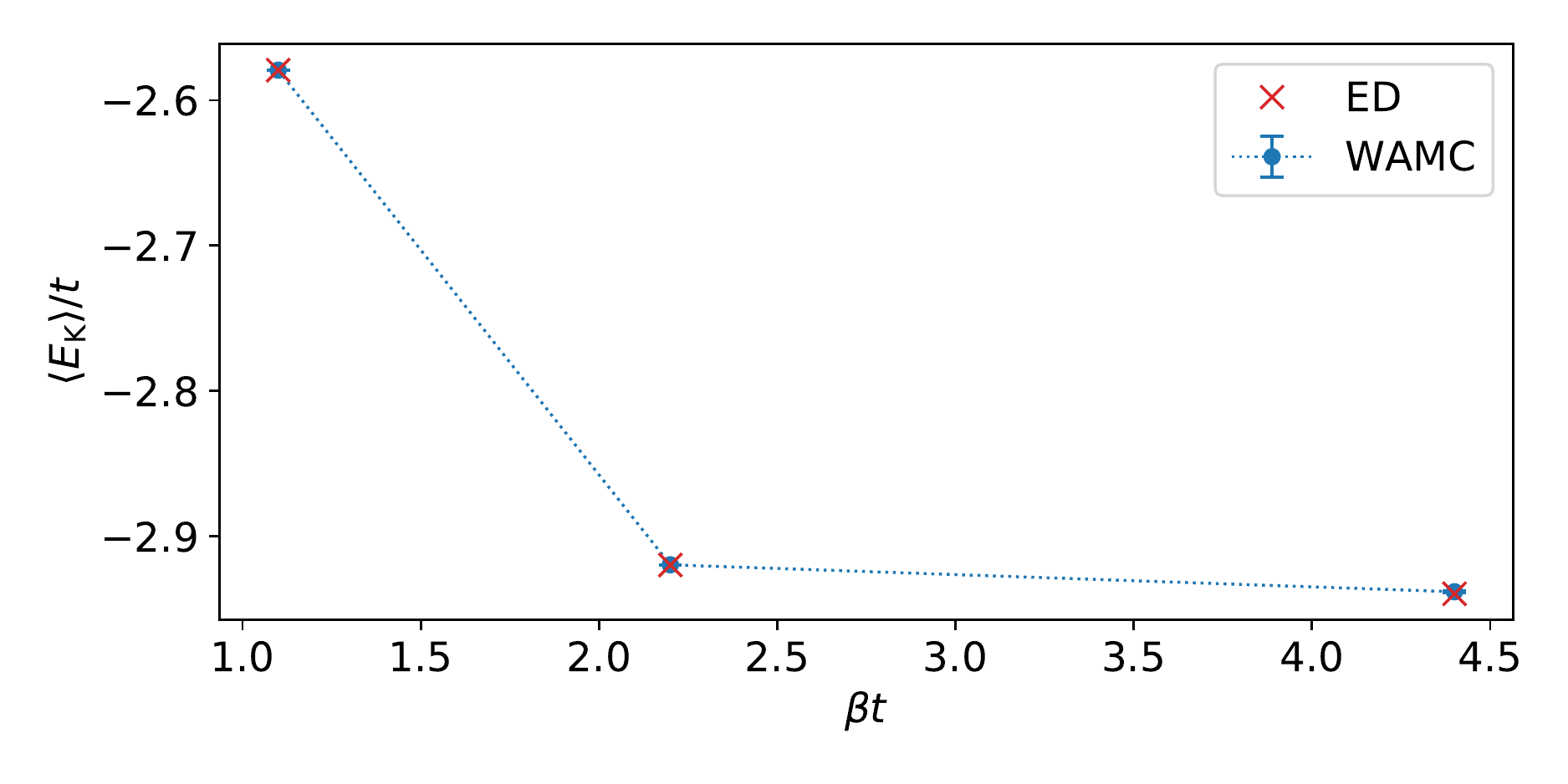}
  \caption{
    Kinetic energy of the polaron obtained via WAMC and ED, as a function of inverse temperature, for a system with $ 4 \times 4 $ lattice sites and periodic boundary conditions. Parameter values are $ J/t = 0.3 $. We find excellent agreement between the data obtained via WAMC and ED. While error bars are given for the WAMC data, they are smaller than the marker size.
  }
  \label{fig:energy_exact_vs_MC}
\end{figure}

\vspace{0.5em}
\textbf{Exact diagonalization.} In order to verify the accuracy of WAMC we benchmarked results for a $ 4 \times 4 $ system against exact diagonalization (ED) \cite{doi:10.1063/1.3518900}.
%
For the correlators $ C_1 $, $ C_{ \sqrt 2 } $, and $ C_2 $, we found perfect agreement, as shown in Fig.\ \ref{fig:CX_exact_vs_MC}. These correlators are defined through
\begin{equation}
  C_{|\mbf d|}(\mbf r) = 4 \langle S^z_{\mbf r_0 + \mbf r - \mbf d / 2} \, S^z_{\mbf r_0 + \mbf r + \mbf d / 2} \rangle \,,
\end{equation}
where $ \mbf r_0 $ and $ \mbf r_0 + \mbf r \pm \mbf d / 2 $ are the positions of the carrier and the two spins, respectively. See Fig.\ 1 of the main text for an illustration. We also found the kinetic energy to perfectly match the exact diagonalization data, which is shown in Fig.\ \ref{fig:energy_exact_vs_MC}.

\bibliography{biblio}